\documentclass[11pt]{article}
\usepackage{amsmath, amssymb, amsfonts, amsthm}

\newcommand{\rd}{{\rm d}}

\newcommand{\bx}{{\bf x}}

\newcommand{\ph}{\varphi}

\newcommand{\bR}{{\mathbb R}}

\newcommand{\bN}{{\mathbb N}}

\newcommand{\tr}{\mbox{Tr}}

\newcommand{\cE}{{\cal E}}

\newcommand{\cH}{{\cal H}}

\newtheorem{theorem}{Theorem}[section]

\begin{document}

\title{Effective Evolution Equations from \\ Many Body Quantum Dynamics}

\author{Benjamin Schlein \\ \\ 
Institute for Applied Mathematics, University of Bonn \\ 
Endenicher Allee 60, 53115 Bonn, Germany \\ 
\texttt{benjamin.schlein@hcm.uni-bonn.de}}

\maketitle

\begin{abstract}
In these notes we review some recent results concerning the derivation of effective equations from first principle quantum dynamics. In particular, we discuss the derivation of the semi-relativistic Hartree equation for the evolution of boson stars, and the derivation of the Gross-Pitaevskii equation for the dynamics of Bose-Einstein condensates.
\end{abstract}

\section{Introduction}
\label{sec:intro}

Systems of interest in physics are typically composed by a huge number of elementary particles. Dilute samples of Bose-Einstein condensates contain $10^3-10^6$ atoms (and, strictly speaking, each atom contains many elementary components). The number of molecules contained in chemical samples is typically of the order of Avogadro's number, $N_A \simeq 6 \cdot 10^{23}$. Systems of relevance in astronomy and cosmology (like stars and galaxies) are composed by up to $10^{60}$ elementary components.

\medskip

In principle, the dynamics of these systems can be determined by solving fundamental evolution equations like the Newton equation or the many-body Schr\"odinger equation. In practice, however, fundamental equations are impossible to solve (neither analytically nor numerically) when so many particles are involved (unless the interaction among the particles is neglected). Moreover, observers are not interested in determining the precise evolution of every particle. Instead, they need a prediction for the macroscopically measurable properties of the dynamics (which result by averaging over the many particles in the system). For this reason, it is very important to find effective evolution equations which, on the one hand, can be easily solved (numerically), and, on the other hand, accurately predict the macroscopic behavior of the system under consideration. One of the main goal of statistical mechanics consists therefore in the development of effective theories approximating the solutions of fundamental evolution equations in the relevant regimes. In these notes, we are going to discuss two examples of systems of interest in physics, for which the derivation of effective evolution equations can be made rigorous in a mathematical sense. In Section \ref{sec:bosonstar}, we will illustrate the derivation of a semi-relativistic Hartree equation for the evolution of boson stars. In Section~\ref{sec:GP}, we will sketch the derivation of the Gross-Pitaevskii equation for the dynamics of initially trapped Bose-Einstein condensates. In both cases, the starting point of our analysis is the fundamental many-body Schr\"odinger equation for bosonic systems. In the rest of the introduction, we show in an abstract setting how effective evolution equations emerge from many body quantum dynamics in certain regimes.  


\medskip

We consider quantum mechanical systems of $N$ spinless bosons in three dimensions (the spin does not play any role, and therefore will be neglected). We describe these systems on the Hilbert space $\cH_N = L^2_s (\bR^{3N})$, consisting of all functions in $L^2 ({\bR}^{3N})$ which are symmetric with respect to arbitrary permutations of the $N$ particles. The time evolution is then described by the $N$ particle Schr\"odinger equation
\begin{equation}\label{eq:schr} i\partial_t \psi_{N,t} = H_N \psi_{N,t} \end{equation} for the wave function $\psi_{N,t} \in {\cal H}_N$. We consider Hamilton operators of the form \begin{equation} H_N = \sum_{j=1}^N  -\Delta_{x_j} + \lambda \sum_{i<j}^N V (x_i - x_j) \, , \end{equation} where $\lambda \in \bR$ is a coupling constant and where the potential $V(x)$ describes the (two-body) interaction (the precise form of $V$ depends on the system under consideration).  Here and in the following, we choose units so that Planck's constant $\hbar = 1$ and the mass of the particles $m = 1/2$. We restrict our attention to (approximately) factorized initial data, where all particles are essentially described by the same orbital. The kinetic energy is then of the order $N$, while the potential energy is of the order $\lambda N^2$. To obtain a non-trivial effective evolution equation in the limit of large $N$, we have to assume that $\lambda N^2 \simeq N$, hence that $\kappa := \lambda N$ is a quantity of order one. This regime is known as the mean-field regime. 

\medskip

To analyze the mean-field regime, consider, at time $t=0$, the factorized initial data $\psi_{N,t=0} = \ph^{\otimes N}$, for an arbitrary one-particle orbital $\ph \in L^2 (\bR^3)$. 
Because of the interaction, factorization is not preserved by the time evolution. However, if $N \gg 1$ and $\kappa = N \lambda$ is of order one, the interaction is very weak and one may still expect factorization to be approximately (and in an appropriate sense) preserved: 
\begin{equation}\label{eq:approx} \psi_{N,t} \simeq \prod_{j=1}^N \ph_t (x_j) \end{equation}
for an appropriate evolved one-particle orbital $\ph_t \in L^2 (\bR^3)$. If this is true, it is very easy to derive the self-consistent nonlinear Hartree equation 
\begin{equation}\label{eq:hartree} i\partial_t \ph_t = -\Delta \ph_t + \kappa (V*|\ph_t|^2) \ph_t \end{equation}
for the one-particle orbital $\ph_t$. This simple argument suggests that the Hartree equation (\ref{eq:hartree}) gives an effective description of the evolution of initially factorized bosonic systems in the mean-field regime characterized by $N \ll 1$ and fixed $\kappa := N \lambda$. 


\medskip

To obtain a mathematical precise statement, we need to specify in which sense  (\ref{eq:approx}) holds true. To this end, we define, for $k =1, \dots ,N$, the $k$-particle reduced density matrix associated with $\psi_{N,t}$ by taking the partial trace  \[ \gamma^{(k)}_{N,t} = \tr_{k+1, \dots ,N} \, |\psi_{N,t} \rangle \langle \psi_{N,t}| \]
where $|\psi_{N,t}\rangle \langle \psi_{N,t}|$ denotes the orthogonal projection onto $\psi_{N,t}$. In other words, $\gamma^{(k)}_{N,t}$ is defined as a non-negative trace-class operator on $L^2 (\bR^{3k})$ with kernel given by
\[ \gamma_{N,t}^{(k)} ( \bx_k ; \bx'_k) = \int d\bx_{N-k} \, \psi_{N,t} (\bx_k , \bx_{N-k}) \overline{\psi}_{N,t} (\bx'_k , \bx_{N-k}) \, ,   \]
where $\bx_k = (x_1, \dots ,x_k)$, $\bx'_k = (x'_1, \dots , x'_k)$, $\bx_{N-k} = (x_{k+1}, \dots , x_{N})$. Note the normalization $\tr \, \gamma^{(1)}_{N,t} = 1$. 

\medskip

The next theorem, which holds under suitable assumptions on the potential $V$, tells us that (\ref{eq:approx}) can be understood as convergence (in the limit of large $N$) of the reduced densities associated to $\psi_{N,t}$. 

\begin{theorem}\label{thm:mf}
Let $\ph \in H^1 (\bR^3)$, $\| \ph \| = 1$, $\kappa\in \bR$ and let $\psi_{N,t} = e^{-iH_N t} \ph^{\otimes N}$ be the solution of the Schr\"odinger equation (\ref{eq:schr}), with initial data $\psi_N = \ph^{\otimes N}$ and with Hamilton operator
\begin{equation} H_N = \sum_{j=1}^N  -\Delta_{x_j} + \frac{\kappa}{N} \sum_{i<j}^N V (x_i - x_j) \, . \end{equation}
Then, if $\gamma^{(k)}_{N,t}$ denotes the $k$-particle reduced density associated with $\psi_{N,t}$, we have, for every fixed $k\in \bN$ and $t \in \bR$, 
\[ \gamma_{N,t}^{(k)} \to |\ph_t \rangle \langle \ph_t|^{\otimes k} \] 
as $N \to \infty$. Here, the convergence is in the trace norm and $\ph_t$ is the solution of the Hartree equation (\ref{eq:hartree}) with $\ph_{t=0}= \ph$. 
\end{theorem}

%

The first proof of Theorem \ref{thm:mf} has been obtained by Spohn in \cite{Sp} for bounded potentials. In \cite{EY}, Erd\H os and Yau proved Theorem \ref{thm:mf} for the (attractive or repulsive) Coulomb potential $V(x) = \pm 1/|x|$. In \cite{RS}, a joint work with I. Rodnianski, we considered again the Coulomb interaction, but this time we obtained bounds on the rate of the convergence. In \cite{KP}, Knowles and Pickl extend the theorem to more singular potentials (with control of the rate of convergence).

\section{Dynamics of Boson Stars}
\label{sec:bosonstar}

In this section we consider systems of gravitating bosons forming so called boson stars. 
We describe boson stars with the Hamilton operator 
\begin{equation}\label{eq:ham-grav} H_{\mathrm{grav}} = \sum_{j=1}^N \sqrt{1-\Delta_{x_j}} - G \sum_{i<j}^N \frac{1}{|x_i -x_j|} \end{equation}
acting on the Hilbert space $\cH_N = L^2_s (\bR^{3N})$ (we use a relativistic dispersion for the particles, but the interaction is classical). As explained in Section \ref{sec:intro}, we are interested in the dynamics generated by (\ref{eq:ham-grav}) for large $N$ and small $G$, with $NG$ of order one. Since the physical value of the gravitational constant, in our units, is approximatively given by $G_{\mathrm{phys}} \simeq 10^{-40}$ (for bosons with mass comparable to a hydrogen atom), this model can be used to describe boson stars with $N \simeq 10^{40}$ particles. For such values of $N$, it makes sense to fix $\lambda := N G$ and to study the dynamics generated by 
\begin{equation}\label{eq:HNbstar} H_N = \sum_{j=1}^N \sqrt{1-\Delta_{x_j}} - \frac{\lambda}{N} \sum_{i<j}^N \frac{1}{|x_i -x_j|} \end{equation}
in the limit $N \to \infty$. 

\medskip

The Hamiltonian (\ref{eq:HNbstar}) is critical in the following sense. For every $N \in \bN$, there exists a critical coupling constant $\lambda_{\mathrm{crit}} (N)$ such that $H_N$ is bounded below for all $\lambda \leq \lambda_{\mathrm{crit}} (N)$ and such that
\[ \inf_{\psi \in L^2 (\bR^{3N})} \frac{\langle \psi, H_N \psi \rangle}{\| \psi \|^2} = - \infty \] for all $\lambda > \lambda_{\mathrm{crit}} (N)$. It was proven in \cite{LY} that, as $N \to \infty$, $\lambda_{\mathrm{crit}} (N) \to \lambda_{\text{crit}}^H$, where $\lambda_{\text{crit}}^H$ is the critical constant for the Hartree energy 
\begin{equation}\label{eq:enhar} \cE_{\mathrm{Hartree}} (\ph) = \int \rd x \, \left| (1-\Delta)^{1/4} \ph (x)\right|^2 - \frac{\lambda}{2} \int \rd x \rd y \frac{|\ph (x)|^2 |\ph (y)|^2}{|x-y|}\, . \end{equation}
In other words, $\lambda_{\text{crit}}^H$ is such that $\cE_{\mathrm{Hartree}} (\ph) \geq 0$ for all $\ph \in H^{1/2} (\bR^3)$ if $\lambda \leq \lambda^\mathrm{H}_{\mathrm{crit}}$ while, if $\lambda > \lambda^\mathrm{H}_{\mathrm{crit}}$,  \[ \inf_{\ph \in H^{1/2} (\bR^3) , \| \ph \| =1} \cE_{\mathrm{Hartree}} (\ph) = -\infty \, . \] 

\medskip

The criticality of the Hamiltonian (\ref{eq:HNbstar}) is a sign for the instability of boson stars when $\lambda = NG > \lambda_{\mathrm{crit}}^{\mathrm{H}}$. If the number of bosons in the star exceeds the critical value $N_\mathrm{crit} = \lambda_{\mathrm{crit}}^{\mathrm{H}}/G$, the star collapses. 


\medskip

Next, we focus on the properties of the evolution generated by the Hamiltonian $H_N$ defined in (\ref{eq:HNbstar}). For the subcritical regime, we show in \cite{ES}, a joint work with A. Elgart, that the many body dynamics can be approximated (in the sense of Theorem \ref{thm:mf}) by the solution of the relativistic Hartree equation 
\begin{equation}\label{eq:hartree-rel}  i\partial_t \ph_t = \sqrt{1-\Delta} \,  \ph_t - \lambda \left( \frac{1}{|.|} * |\ph_t|^2 \right) \ph_t  \, . \end{equation} Note that, in the subcritical regime, Lenzmann showed in \cite{L} that Eq. (\ref{eq:hartree-rel}) is globally well-posed in the energy space $H^{1/2} (\bR^3)$. What about the supercritical regime? Since $H_N$ is not bounded from below, it is not a priori clear how to define the one-parameter group of unitary transformations $U_N (t) = e^{-iH_N t}$ describing the time-evolution. To circumvent this problem, we introduce a tiny, $N$-dependent, cutoff $\alpha (N)$ in the Coulomb potential, 
\begin{equation}\label{eq:ham-reg} H_N^\alpha = \sum_{j=1}^N \sqrt{1-\Delta_{x_j}} - \frac{\lambda}{N} \sum_{i<j}^N \frac{1}{|x_i -x_j|+\alpha (N)} \end{equation}
and we assume that $\alpha (N) \to 0$ as $N \to \infty$. The regularized Hamiltonian $H^{\alpha}_N$ is now bounded below for every $N$. Therefore it can be extended (uniquely) to a self-adjoint operator on $\cH_N$, the unitary group $U_N (t) = e^{-iH_N^\alpha t}$ is well defined and the Schr\"odinger equation \begin{equation}\label{eq:schr-reg}
i\partial_t \psi_{N,t} = H_N^{\alpha} \psi_{N,t} \end{equation} is globally well posed on $\cH_N$.  On the other hand, since the cutoff vanishes in the limit $N \to \infty$, we may still expect the effective dynamics to be described by the semi-relativistic Hartree equation (\ref{eq:hartree-rel}).

\medskip

It is important to notice that the criticality of the model can also be observed at the level of (\ref{eq:hartree-rel}). For $\lambda > \lambda^\mathrm{H}_{\mathrm{crit}}$, the equation is still locally well-posed in $H^{1/2} (\bR^3)$ (for an arbitrary initial data $\ph$ in $H^{1/2} (\bR^3)$ there exists a unique solution $\ph_t$ in $H^{1/2}$ on the time interval $t \in (-T,T)$, for some $T>0$). In general, however, the local solution cannot be extended to a global solution (i.e., one cannot take $T=\infty$). 
In fact, for $\lambda  > \lambda^\mathrm{H}_{\mathrm{crit}}$, it was proven by Fr\"ohlich and Lenzmann in \cite{FL} that there exist solutions of (\ref{eq:hartree-rel}) which exhibit blow-up in finite time. This means that there are solutions $\ph_t$ of (\ref{eq:hartree-rel}), and $0< T < \infty$ such that \begin{equation} \| \ph_t \|_{H^{1/2}} = \left(\int dx \left| (1-\Delta)^{1/4} \ph_t (x) \right|^2 \right)^{1/2} \to \infty \end{equation} as $t \to T^-$. Solutions of (\ref{eq:hartree-rel}) exhibiting blow-up in finite time can be used to give a dynamical description of the phenomenon of gravitational collapse. 
 
\medskip

The next two theorems from \cite{MS}, a joint work with A. Michelangeli, show that, also in the supercritical regime, the solution $\ph_t$ to the relativistic Hartree equation (\ref{eq:hartree-rel}) continues to approximate the many body dynamics until the time where $\ph_t$ blows up (if $\ph_t$ does not exhibit blow up, then it stay close to the solution of the many-body Sch\"rodinger equation on every finite time interval). 

\medskip

The first theorem proves that, if $\ph_t$ does not blow up in the time interval $[-T,T]$, $|\ph_t \rangle \langle \ph_t|$ is close to $\gamma^{(1)}_{N,t}$, as $N \to \infty$, for all $|t| < T$. 
\begin{theorem}
Fix $\ph \in H^2 (\bR^3)$ with $\| \ph \| = 1$ and $T>0$ such that 
\begin{equation}\label{eq:ka1}
\kappa := \sup_{|t|\leq T} \|\ph_t\|_{H^{1/2}} <  \infty \,
\end{equation}
where $\ph_t$ is the solution of (\ref{eq:hartree-rel}) with initial data $\ph_{t=0} = \ph$. Let $\psi_{N,t} = e^{-iH^{\alpha}_N t} \ph^{\otimes N}$. Then there exists a constant $C$ (depending only on $T$, $\| \ph \|_{H^2}$, and $\kappa$) such that 
\begin{equation}\label{eq:conv1-tr}
\tr \, \left| \gamma_{N,t}^{(1)} - |\ph_t\rangle\langle\ph_t| \right| \leq \frac{C}{\sqrt{N}}  
\end{equation}
for all $|t| \leq T$.
\end{theorem}

The next theorem shows that if $\ph_t$ blows up at time $T$, then also the solution to the many body Schr\"odinger equation $\psi_{N,t}$ collapses as $t$ approaches $T$, if, at the same time, $N \to \infty$ sufficiently fast. This result justifies the use of the Hartree equation (\ref{eq:hartree}) for the description of the gravitational collapse of boson stars. 
\begin{theorem}\label{thm:blow}
Fix $\ph \in H^2 (\bR^3)$ with $\| \ph \| = 1$. Suppose that $T_c >0$ is the first time of blow-up for the solution $\ph_t$ of (\ref{eq:hartree}) with initial data $\ph$ ($\| \ph_t \|_{H^{1/2}} < \infty$ for all $t \in [0,T_c)$ and $\| \ph_t \|_{H^{1/2}} \to \infty$ as $t \to T_c^{-}$). Let $\psi_{N,t} = e^{-iH^{\alpha}_N t} \ph^{\otimes N}$, and assume that, in (\ref{eq:ham-reg}), $\alpha(N) \geq N^{-\beta}$ for some $\beta >0$. 
Then, for every $|t| < T_c$, there exists a constant $C_t < \infty$ such that 
\[ \| (1-\Delta_{x_1})^{1/4} \, \psi_{N,t} \|  \leq C_t \] uniformly in $N$. Moreover, if $N(t) \in \bN$ for $t \in [0,T_c)$ is so that $N(t) \to \infty$ sufficiently fast as $t \to T_c$, we have  
\begin{equation}
\| (1-\Delta_{x_1})^{1/4} \, \psi_{N(t),t} \|^2 = \tr \, (1-\Delta)^{1/2}\gamma_{N(t),t}^{(1)}\to \infty\qquad\mathrm{as}\quad t\to T^-_c\, .
\end{equation}
\end{theorem}
To show these two theorems we use the techniques developed in \cite{RS}. These techniques were first introduced, in a slightly different context, by Hepp in \cite{He}. They are based on the representation of the many-body system on the Fock space, and on the use of coherent states as initial data. With respect to \cite{RS}, the main novelty is that, to prove Theorem \ref{thm:blow}, we need to show convergence not only in the trace norm (as in (\ref{eq:conv1-tr})), but also in the energy norm. In other words, we have to show that, as long as $\ph_t$ does not blow up, 
\begin{equation}\label{eq:energy-conv} \tr \, \left| (1-\Delta)^{1/4} \left( \gamma^{(1)}_{N,t} - |\ph_t \rangle \langle \ph_t| \right) (1-\Delta)^{1/4} \right| \to 0 \end{equation} as $N \to \infty$. In fact, as pointed out to us by R. Seiringer, the existence of $0 < T_0 \leq T_c$ with the property $\|  (1-\Delta_{x_1})^{1/4} \, \psi_{N(t),t} \|^2 \to \infty$ as $t \to T_0^-$ (with $N(t) \to \infty$ as $t \to T_0^-$) follows from (\ref{eq:conv1-tr}) and from the semicontinuity of the kinetic energy (the kinetic energy of the limit is always smaller than the limit of the kinetic energy). However, it is only (\ref{eq:energy-conv}) that allows us to conclude that 
$T_0 = T_c$ and therefore that the collapse of the many body wave function can really be described by the blow up of the solution of the Hartree equation (\ref{eq:hartree-rel}).

\section{Dynamics of Bose-Einstein condensates}
\label{sec:GP}

Since the work of groups around Cornell and Wieman at the University of Colorado, and around Ketterle at MIT, see \cite{CW,Kett}, Bose-Einstein condensation has become accessible to experiments. In these experiments dilute Bose gases are initially trapped by strong magnetic fields. Then, after cooling the gas to very low temperatures (of the order of nano-kelvin), the traps are switched off and the evolution of the gas is observed. 
To understand these experiments it is important to find an accurate description of the macroscopic properties of the evolution of the condensate.

\medskip

The trapped Bose gas is described by the Hamiltonian 
\begin{equation}
H_N^{\mathrm{trap}} = \sum_{j=1}^N  (-\Delta_{x_j} + V_{\mathrm{ext}} (x_j) ) + \sum_{i<j}^N N^2 V (N (x_i -x_j)) 
\end{equation}
acting on the $N$ boson Hilbert space $\cH_N = L^2_s (\bR^{3N})$. Here $V_{\mathrm{ext}}$ is an external potential modeling the magnetic traps, $V \geq 0$ is repulsive and of short range, and the interaction potential $V_N (x) = N^2 V (Nx)$ scales with the number of particles $N$ so that its scattering length is of the order $1/N$. Recall  that the scattering length of $V$ is defined as 
\begin{equation}\label{eq:scattlg} 8\pi a_0 = \int dx \, V(x) f(x)\, \end{equation}
where $f$ is the solution of the zero-energy scattering equation \begin{equation}\label{eq:scatt} \left( -\Delta + \frac{1}{2} V \right) f = 0 \end{equation} with the boundary condition $f (x) \to 1$ as $|x| \to \infty$ (it is then simple to check that $f(x) \simeq 1 - (a_0/|x|) + O(|x|^{-2})$ as $|x| \to \infty$). If $f$ is a solution of (\ref{eq:scatt}), it follows by scaling that $f_N (x) = f (Nx)$ solves \begin{equation}\label{eq:zeroN} \left( -\Delta + \frac{1}{2} V_N \right) f_N = 0 \,. \end{equation} This implies, in particular, that the scattering length of $V_N$ equals $a_0/N$.

\medskip

In \cite{LSY}, Lieb, Seiringer and Yngvason proved that, if $E_N$ denotes the ground state energy of $H_N$, $E_N/N \to \min_{\ph:\| \ph \|=1} \cE_{GP} (\ph)$ as $N \to \infty$. Here 
\[ \cE_{GP} (\ph) = \int dx \, \left( |\nabla \ph (x)|^2 + V_{\mathrm{ext}} (x)|\ph (x)|^2 + 4 \pi a_0 |\ph (x)|^4 \right) \]
for all $\ph \in L^2 (\bR^3)$ is the so called Gross-Pitaevskii energy functional. 
In \cite{LS}, Lieb and Seiringer showed that the ground state of $H_N^{\mathrm{trap}}$ exhibits complete condensation into the minimizer $\phi_{GP}$ of $\cE_{GP}$, in the sense that \begin{equation}\label{eq:cond} \gamma_N^{(1)} \to |\phi_{GP} \rangle \langle \phi_{GP}| \qquad \text{as } \quad N \to \infty\,, \end{equation}
where $\gamma_N^{(1)}$ is the one-particle density of the ground state of $H_N^{\mathrm{trap}}$.

\medskip

What happens now when the traps are turned off? The system starts to evolve with respect to the translation invariant Hamitonian \begin{equation}\label{eq:HNtr} H_N = \sum_{j=1}^N -\Delta_{x_j} + \sum_{i<j}^N N^2 V (N (x_i -x_j)) \, .\end{equation}
It turns out that the macroscopic properties of the resulting evolution can also 
be approximated by the same Gross-Pitaevskii which successfully describes the properties of the ground state. The following theorem is proven in \cite{ESY1,ESY2,ESY4}, a series of joint works with L. Erd\H os and H.-T. Yau. 

\begin{theorem}\label{thm:GP}
Suppose that $0 \leq V(x) \leq C (1 + x^2)^{-\sigma/2}$ for some $\sigma >5$. 
Let $\psi_N \in L^2 (\bR^{3N})$ be a sequence of $N$-particle wave functions with $\| \psi_N \| = 1$, so that, as $N \to \infty$, $\gamma^{(1)}_N \to |\ph \rangle \langle \ph|$ for an arbitrary $\ph \in H^1 (\bR^3)$ ($\psi_N$ exhibits complete condensation) and so that $\langle \psi_N , H_N \psi_N \rangle \leq C N$ ($\psi_N$ has finite energy per particle). Then, for every fixed $t \in \bR$, the evolved wave function $\psi_{N,t} = e^{-iH_N t} \psi_N$ still exhibits complete Bose-Einstein condensation, in the sense that 
\begin{equation}\label{eq:convGP} \gamma_{N,t}^{(1)} \to |\ph_t \rangle \langle \ph_t| \end{equation} as $N \to \infty$. Here $\ph_t$ is the solution of the Gross-Pitaevskii equation \begin{equation}\label{eq:GP} i\partial_t \ph_t = -\Delta \ph_t + 8 \pi a_0 |\ph_t|^2 \ph_t \end{equation} with the initial data $\ph_{t=0} = \ph$. 
\end{theorem}

Observe that (\ref{eq:convGP}) immediately implies that $\gamma^{(k)}_{N,t} \to |\ph_t \rangle \langle \ph_t|^{\otimes k}$ for all $k \geq 1$ (this is a general properties of densities converging to rank one projections). This theorem shows that  the Gross-Pitaevskii equation can be used to describe the evolution of the condensates in the experiments we discussed above. 
In the rest of this section, I will explain the main ideas of the proof of Theorem~\ref{thm:GP}. We follow the general strategy introduced in \cite{Sp} to analyze the mean field limit; we need however to adapt the techniques to the present situation. The main difficulties are a consequence of the fact that (\ref{eq:HNtr}) does not describe a mean field regime, which is characterized by many weak collisions among the particles. On the contrary, the evolution generated by (\ref{eq:HNtr}) is characterized by very rare and very strong collisions (particles interact only when they are extremely close, at distances of order $1/N$).

\medskip

The starting point of our analysis is the study of the time evolution of the reduced density matrices associated with the solution of the Schr\"odinger equation $\psi_{N,t}$. It turns out that the family $\gamma^{(k)}_{N,t}$, for $k=1, \dots ,N$, satisfies a hierarchy of $N$ coupled equations, commonly known as the BBGKY hierarchy:
\begin{equation}\label{eq:BBGKY}
\begin{split}
i\partial_t \gamma^{(k)}_{N,t} = \; & \sum_{j=1}^k \left[ -\Delta_{x_j} , \gamma^{(k)}_{N,t} \right] + \sum_{i<j}^k \left[ N^2 V (N (x_i -x_j)) , \gamma^{(k)}_{N,t} \right] \\ &+ (N-k) \sum_{j=1}^k \tr_{k+1} \, \left[ N^2 V (N (x_j - x_{k+1})) , \gamma^{(k+1)}_{N,t} \right] 
\end{split}
\end{equation}
where $\tr_{k+1}$ denotes the partial trace over the degrees of freedom of the $(k+1)$-th particle. {F}rom this hierarchy, we try to find equations for the limit points of the reduced densities, as $N \to \infty$. Suppose, for example, that $\gamma^{(1)}_{\infty,t}$ and $\gamma^{(2)}_{\infty,t}$ are limit points of $\gamma^{(1)}_{N,t}$ and, respectively, of $\gamma^{(2)}_{N,t}$. Then, from (\ref{eq:BBGKY}), with $k=1$, 
we may expect $\gamma^{(1)}_{\infty,t}$ and $\gamma^{(2)}_{\infty,t}$ to satisfy the equation 
\begin{equation}\label{eq:wrong} i\partial_t \gamma^{(1)}_{\infty,t} = [-\Delta, \gamma^{(1)}_{\infty,t}] + \tr_2 \left[ b_0 \delta (x_1 -x_2), \gamma^{(2)}_{\infty,t} \right] \end{equation}
where we used that, as $N \to \infty$, \[ (N-1) N^2 V (N (x_1 -x_2)) \to b_0 \delta (x_1 - x_2), \quad \text{with } \quad b_0 = \int V(x) dx.\] It turns out that (\ref{eq:wrong}) is not correct. Taking the limit $N \to \infty$ in (\ref{eq:BBGKY}), we ignored the correlations developed by the two point reduced density $\gamma^{(2)}_{N,t}$ on the length scale $1/N$. Using the solution $f_N (x) = f (Nx)$ of the zero energy scattering equation (\ref{eq:zeroN}) to describe the correlations, we can try to approximate $\gamma_{N,t}^{(1)}$, $\gamma^{(2)}_{N,t}$, for large but finite $N$, by  
\begin{equation} 
\begin{split}
\gamma^{(1)}_{N,t} (x_1 ; x'_1) &\simeq \gamma^{(1)}_{\infty,t} (x_1; x'_1) , \\
\gamma^{(2)}_{N,t} (x_1, x_2 ; x'_1, x'_2) &\simeq f_N (x_1 -x_2) f_N (x'_1 - x'_2) \gamma^{(2)}_{\infty,t} (x_1, x_2 ; x'_1, x'_2) \,.
\end{split}
\end{equation}
Inserting this ansatz in (\ref{eq:BBGKY}) (for $k=1$), and using the definition (\ref{eq:scattlg}) of the scattering length, we easily find another (and, this time, correct) equation for the limit points $\gamma^{(1)}_{\infty,t}, \gamma^{(2)}_{\infty,t}$:
\begin{equation}\label{eq:gm1gm2} i\partial_t \gamma^{(1)}_{\infty,t} = \left[ -\Delta, \gamma^{(1)}_{\infty,t} \right] + 8 \pi a_0 \tr_2 \, \left[ \delta (x_1 -x_2) , \gamma^{(2)}_{\infty,t} \right] \, .  \end{equation}
Note that, $f_N \to 1$ as $N \to \infty$ (in a weak sense). For this reason, $f_N$ only plays an important role when it is coupled with the (very) singular potential $V_N$. 
Similarly to (\ref{eq:gm1gm2}), starting from (\ref{eq:BBGKY}) for $k > 1$ and considering the limit $N \to \infty$ (taking into account the short scale correlation structure developed by $\gamma^{(k+1)}_{N,t}$), we conclude that an arbitrary limit point $\{ \gamma^{(k)}_{\infty,t} \}_{k \geq 1}$ of the sequence of reduced density matrices $\{ \gamma^{(k)}_{N,t} \}_{k=1}^N$ satisfies the infinite hierarchy of equations
\begin{equation}\label{eq:GPhier}
 i\partial_t \gamma^{(k)}_{\infty,t} = \sum_{j=1}^k \left[ -\Delta_{x_j}, \gamma^{(k)}_{\infty,t} \right] + 8 \pi a_0 \sum_{j=1}^k \tr_{k+1} \, \left[ \delta (x_j -x_{k+1}) , \gamma^{(k+1)}_{\infty,t} \right] \, .
\end{equation}
It is then useful to observe that the factorized densities \begin{equation}\label{eq:factor-sol} \gamma^{(k)}_{\infty,t} = |\ph_t \rangle \langle \ph_t|^{\otimes k} \end{equation}
solve (\ref{eq:GPhier}) if and only if $\ph_t$ is a solution of the Gross-Pitaevskii equation (\ref{eq:GP}).

\medskip

The strategy to show Theorem \ref{thm:GP} consists therefore of the following steps. 
\begin{itemize}
\item[1)] Show the compactness of the sequence of families $\{ \gamma^{(k)}_{N,t} \}_{k=1}^N$ with respect to an appropriate weak topology.
\item[2)] Prove that the reduced densities $\gamma^{(k+1)}_{N,t}$, $k \geq 1$, have a short scale correlation structure which can be described, in good approximation, by the solution $f_N$ of the zero energy scattering equation (\ref{eq:zeroN}).
\item[3)] Using the short scale structure developed by $\gamma^{(k+1)}_{N,t}$, show that any limit point of the sequence $\{ \gamma^{(k)}_{N,t} \}_{k=1}^N$ is a solution of the infinite hierarchy (\ref{eq:GPhier}).
\item[4)] Prove the uniqueness of the solution of the infinite hierarchy (\ref{eq:GPhier}).
\end{itemize}
Steps 1-4 conclude the proof of Theorem \ref{thm:GP} because a compact sequence with at most one limit point must be convergent and therefore (\ref{eq:factor-sol}) must be its limit. Details can be found in \cite{ESY1,ESY2,ESY3,ESY4}. Recently, an alternative proof of Theorem~\ref{thm:GP} was proposed by Pickl in \cite{P1,P2}. Pickl's approach also allows for the presence of (possibly time-dependent) external potentials in the Hamiltonian (\ref{eq:HNtr})  (but requires stronger conditions on the initial wave function).  

\end{document}